\documentclass[aps,prd,epsf]{revtex4}
\usepackage[latin1]{inputenc}
\usepackage{float}
\usepackage{graphicx}

\newcommand{\be}{\begin{equation}}
\newcommand{\ee}{\end{equation}}
\newcommand{\bea}{\begin{eqnarray}}
\newcommand{\eea}{\end{eqnarray}}

\def\ty_3{\tilde{y}_3}

\newcommand{\rd}{{\rm d}}

\begin{document}

\title{{\large UV stable, Lorentz-violating dark energy 
with transient phantom era}}


\author{Maxim Libanov}
\affiliation{Institute for Nuclear Research of the Russian Academy of
Sciences,\\
60th October Anniversary Prospect, 7a, Moscow, 117312, Russia}

\author{Eleftherios Papantonopoulos}
\affiliation{Department of Physics, National Technical University of
Athens,\\
Zografou Campus GR 157 73, Athens, Greece}

\author{Valery Rubakov} \affiliation{Institute for Nuclear Research of the
Russian Academy of Sciences,\\
60th October Anniversary Prospect, 7a, Moscow, 117312, Russia}

\author{M.~Sami}
\affiliation{Centre for Theoretical Physics, Jamia Millia, New
Delhi-110025, India}

\author{Shinji Tsujikawa}
\affiliation{Department of Physics, Gunma National College of
Technology, Gunma 371-8530, Japan}

\begin{abstract}

Phantom fields with negative kinetic energy are  often plagued by
the vacuum quantum instability in the ultraviolet region. We
present a Lorentz-violating dark energy model free from this problem and
show that the crossing of the cosmological constant boundary $w=-1$ to the
phantom equation of state is realized before reaching a de~Sitter
attractor. Another  interesting feature is a peculiar
time-dependence of the effective Newton's constant; the magnitude of this
effect is naturally small but may be close to experimental limits. We also
derive momentum scales of instabilities at which tachyons or ghosts appear
in the infrared region around the present Hubble scale and clarify the
conditions under which tachyonic instabilities do not spoil homogeneity of
the present/future Universe.

\end{abstract}

\maketitle

\section{Introduction}

The compilations of various observational data show that the Universe has
entered the stage of an accelerated expansion around the redshift $z \sim
1$ \cite{SN,WMAP,LSS,BAO,review,CST}. The equation of state (EOS)
parameter $w$ of Dark Energy (DE) responsible for the acceleration of the
Universe has been constrained to be close to $w=-1$. However, the phantom
EOS ($w<-1$) is still allowed by observations and even favored by some
analyses of the data \cite{phanobser}. It is also possible that the EOS of
DE crossed the cosmological constant boundary ($w=-1$) in relatively near
past~\cite{crossing}.

The presence of the phantom corresponds to the violation of weak energy
condition, the property which is generally difficult to accommodate within
the framework of field theory. The simplest model which realizes the
phantom EOS is provided by a minimally coupled scalar field with a
negative kinetic term \cite{phanori,Carroll} (see also
Refs.~\cite{phanpapers,quintom}). The negative kinetic energy is generally
problematic because it leads to a quantum instability of the vacuum in the
ultraviolet (UV) region
\cite{Carroll,Cline,ghostcondensate,dghostcondensate,Hsu,Dub}: the vacuum
is unstable against the catastrophic particle production of ghosts and
normal (positive energy) fields.

There have been a number of attempts to realize the phantom EOS without
having the pathological behaviour in the UV region. One example is
scalar-tensor gravity in which a scalar field $\phi$ with a positive
kinetic term is coupled to Ricci scalar $R$ \cite{scatensor,Boi}. This
coupling leads to the modification of gravitational constant, but it was
shown in Ref.~\cite{Gano} that there are some parameter regions in which a
phantom effective EOS is achieved without violating local gravity
constraints in the present Universe.

Another example is provided by the so-called modified gravity, including
$f(R)$ gravity models \cite{fR} and the Gauss-Bonnet (GB) models
\cite{GB}. In $f(R)$  models  it is possible to obtain a strongly phantom
effective EOS, but in that case the preceding matter epoch is practically
absent \cite{Gano2}. For GB DE models it was shown in Ref.~\cite{Mota}
that the crossing of the cosmological constant boundary, $w=-1$, is
possible, but local gravity experiments place rather strong constraints on
the effective GB energy fraction \cite{LucaGB}. In addition, tensor
perturbations are typically plagued by instabilities in the UV region if
the GB term is responsible for the accelerated expansion of the Universe
 \cite{ten}. Thus, it is generally not so easy to construct viable
modified gravity models that realize the phantom effective EOS  without
violating cosmological and local gravity constraints.

The third example is the Dvali-Gabadadze-Porrati (DGP) braneworld
model~\cite{DGP} and its extension~\cite{kofi} with a GB term in the bulk,
which allow for the possibility to have $w<-1$ \cite{Sahni,Lue}. However,
it was shown in Ref.~\cite{Koyama} that the DGP model contains a ghost
mode, which casts doubts on the viability of the self-accelerating
solution.

While the above models more or less correspond to the modification of
gravity, it was recently shown that in the Einstein gravity in a
Lorentz-violating background the phantom  EOS can be achieved without any
inconsistency in the UV region \cite{Crem2,Rubakov}. In particular, in the
model of Ref.~\cite{Rubakov} Lorentz invariance is broken in the presence
of a vector field $B_{\mu}$ which has two-derivative kinetic terms similar
to those given in Ref.~\cite{Ben}. The effect of the Lorentz violation is
quantified by a parameter $\Xi \equiv B_{\mu}B^{\mu}/{\cal M}^2$, where
${\cal M}$ is an UV cut-off scale. In analogy to Ref.~\cite{MLVR} the
vector field also has one-derivative coupling $\epsilon \partial_{\mu}\Phi
B^{\mu}$ with a scalar field $\Phi$, where $\epsilon$ is a small parameter
that characterizes an IR scale. In the UV region, where the spatial
momentum $p$ is much larger than $\epsilon$,  ghosts, tachyons and
super-luminal modes are not present. Meanwhile tachyons or ghosts can
appear in the IR region $p \lesssim \epsilon$. This is not problematic
provided that $\epsilon$ is close to the  present Hubble scale.

In this paper we apply this Lorentz-violating model to dark energy and
study the cosmological dynamics in detail in the presence of mass terms in
the potential, $V=\frac12 m^2\Phi^2-\frac12 M^2 X^2$ (where 
$X^2=B_\mu B^\mu$). We
show that the model has a de~Sitter attractor responsible for the
late-time acceleration. At early times DE naturally has normal EOS with
 $w>-1$, while the phantom EOS can be realized between the
matter-dominated era and the final de~Sitter epoch. We clarify the
conditions under which the cosmological constant boundary crossing to the
phantom region occurs. Interestingly, in a range of parameters this
crossing takes place at the epoch when $\Omega_m\sim\Omega_{\rm DE}$ thus
making the crossing potentially observable.

 Another interesting feature of our model is the time-dependence of the
effective Newton's constant.
It is naturally weak, but may well be
comparable with current experimental limits. Moreover, the effective
Newton's constant $G_{*}(t)$ has a peculiar behaviour correlated with the
deviation of $w$ from $-1$.

We  also derive momentum scales of instabilities of perturbations, first
in Minkowski spacetime. This is the extension of the work \cite{Rubakov}
that mainly focused on the case of massless scalar ($m=0$). We show that
in the UV region ($p \gg \epsilon$) the model does not have any unhealthy
states such as ghosts, tachyons or super-luminal modes. In the IR region
($p \lesssim \epsilon$) tachyons or ghosts appear, depending on the
momentum. Finally, we study the evolution of perturbations in the
cosmological background and estimate the amplitude of perturbations
amplified by the tachyonic instability around the scale of the present
Hubble radius. The perturbations remain to be smaller than the background
fields under certain restriction on the model parameters.

This paper is organized as follows. In Sec.~\ref{modelsec} we present our
Lorentz-violating model and derive basic equations  describing spatially
flat Friedmann--Robertson--Walker cosmology in the presence of DE, radiation
and non-relativistic matter. In Sec.~\ref{dynamicssec}  the
cosmological dynamics is discussed in detail analytically and numerically
with an emphasis on the occurrence of a phantom phase before reaching a
de~Sitter attractor.  The time-dependence of the effective gravitational
constant is also considered.  In Sec.~\ref{instabilitysec} we study 
the Minkowski spectrum of field
perturbations and clarify the properties of tachyons and ghosts in the IR
region. We then discuss the tachyonic amplification of field perturbations
around the present Hubble scale in the cosmological background. We
summarize our results in Sec.~\ref{conclusionsec}.
Appendix A contains the derivation of the effective ``Newtonian
gravitational constant'' in our model. In Appendix B we derive the
fixed points of the system by rewriting the equations in autonomous
form. We analyse the stability of the fixed points and show
analytically that the cosmological evolution proceeds from radiation-dominated
stage through matter-dominated stage to the final de~Sitter regime.

\section{Lorentz-violating model}
\label{modelsec}

We study a 4-dimensional Lorentz-violating model whose Lagrangian density
includes a vector field $B_\mu$ and a scalar field $\Phi$:
\begin{eqnarray}
\label{lag}
{\cal L}=-\frac12 \alpha (\Xi) g^{\nu \lambda} D_\mu B_\nu
D^\mu B_\lambda + \frac{1}{2}\beta(\Xi) D_\mu B_\nu D^\mu B_\lambda
\frac{B^\nu B^\lambda}{{\cal M}^2}
+\frac12 \partial_{\mu} \Phi \partial^{\mu} \Phi +\epsilon
\partial_\mu \Phi B^\mu-V(B, \Phi) \,,
\end{eqnarray}
where $\Xi=B_\mu B^{\mu}/{\cal M}^2$ with ${\cal M}$ being an UV cut-off
scale of the effective theory. The dimensionless parameters $\alpha$ and
$\beta$ are the functions of $\Xi$, and $\epsilon$ is a free positive
parameter that characterizes an IR scale. The first two terms in
(\ref{lag}) are familiar in two-derivative theory \cite{Ben}, whereas  the
one-derivative term $\epsilon \partial_\mu \Phi B^\mu$ is introduced
following the approach of Ref.~\cite{MLVR}.

We study  dynamics of flat Friedmann-Robertson-Walker (FRW) Universe
\begin{eqnarray}
\label{FRW}
\rd s^2 = {\cal N}^2 (t) \rd t^2 - a^2(t) \rd {\bf x}^2\,,
\end{eqnarray}
where ${\cal N}(t)$ is a Lapse function and $a(t)$ is a scale factor. In
the case of  spatially homogeneous fields with $B_i=0$ ($i=1,2,3$), the
Lagrangian (\ref{lag}) reads
\begin{eqnarray}
\sqrt{-g}{\cal L} = \frac{\gamma}{2} \frac{a^3}{{\cal N}}\dot{X}^2 -
\frac{3\alpha}{2} \frac{\dot{a}^2 a}{{\cal N}} X^2 + \frac{1}{2}
\frac{a^3}{{\cal N}}\dot{\phi}^2 + \epsilon a^3 \dot{\phi} X - a^3{\cal N}
V(X,\phi)\,,
\label{Eqn/Pg5/1:LV16}
\end{eqnarray}
where $X=B_0/{\cal N}$, $\phi$ is the homogeneous part of the field $\Phi$
and
\begin{eqnarray}
\gamma (X)=\frac{X^2}{{\cal M}^2} \beta(X)-\alpha(X)\,.
\end{eqnarray}
Hereafter we study the case in which the following condition holds
\begin{eqnarray}
\nonumber
\alpha>\gamma>0\,.
\end{eqnarray}
This is required to avoid a superluminal propagation in Minkowski
spacetime \cite{Rubakov}, as we will see later. Throughout this paper we
assume that $\alpha$ and $\gamma$ are of order unity.

For fixed $X$, 
the second term in the Lagrangian (\ref{Eqn/Pg5/1:LV16}) 
has precisely the form of the Einstein--Hilbert action specified
to the flat FRW metric. Hence, it
leads
to the change of the ``cosmological'' effective Planck mass \cite{Rubakov}
\begin{eqnarray}
\label{effpla}
m_{{\rm pl, cosm}}^2=m_{\rm pl}^2+ 4\pi \alpha X^2\,.
\end{eqnarray}
Another effective Planck mass $m_{\mathrm{pl, Newton}}$ determines the
strength of gravitational interactions at distances much shorter than
the cosmological scale; in general, these two effective Planck masses are
different~\cite{Boi,CarrollLim, MLVR}. We show in Appendix A that the
``Newtonian'' Planck mass in our model is given by
\begin{eqnarray}
\label{newton}
m_{{\rm pl, Newton}}^2=m_{\rm pl}^2- 4\pi \alpha X^2\,.
\end{eqnarray}
Both effective Planck masses depend on time via $X=X(t)$. Since the
time-dependent terms in (\ref{effpla}) and (\ref{newton}) differ by sign
only, it will be sufficient to study one of these effective masses. In
what follows we concentrate on the ``Newtonian'' mass (\ref{newton}) for
definiteness.

In this paper we focus on the case in which the potential $V$ takes
a separable form:
\begin{eqnarray}
V=W(\phi)+U(X)\,.
\end{eqnarray}
We take into account the contributions of non-relativistic matter and
radiation whose energy densities $\rho_m$ and $\rho_r$, respectively,
satisfy
\begin{eqnarray}
& &\dot{\rho}_m+3H \rho_m=0\,, \label{8*}\\
& &\dot{\rho}_r+4H \rho_r=0\,. \label{9*}
\end{eqnarray}
The energy density of the fields is derived by taking the derivative with
respect to ${\cal N}$ of the action $S=\int {\rm d}^4x \sqrt{-g}{\cal L}$:
\begin{eqnarray}
\rho= -\frac{1}{a^3}\left[ \frac{\delta S}{\delta {\cal N}} \right]_{{\cal
N}=1} = \frac{\gamma}{2} \dot{X}^2 -\frac{3\alpha}{2}H^2 X^2 + \frac{1}{2}
\dot{\phi}^2 + V \,.
\label{page3+}
\end{eqnarray}
We set ${\cal N}=1$ for the rest of this paper.

The Friedmann equation is given by
\begin{eqnarray}
H^2 \equiv \left(\frac{\dot{a}}{a} \right)^2= \frac{\kappa^2}{3}
\left[\frac12 \gamma \dot{X}^2-\frac{3\alpha}{2}H^2 X^2 +\frac12
\dot{\phi}^2+W(\phi)+U(X)+ \rho_m+\rho_r \right]\,,
\label{FR}
\end{eqnarray}
where $\kappa^2=8\pi/m_{\rm pl}^2$. The equations of motion for the
homogeneous fields $\phi$ and $\chi$ are
\begin{eqnarray}
- \gamma \left(\ddot{X}  + 3H\dot{X} \right) - \frac{1}{2}\gamma_{,X}
\dot{X}^2 -\frac{3}{2}\alpha_{,X} H^2 X^2 -3\alpha H^2 X  + \epsilon
\dot{\phi} &=& U_{,X}\,,
\label{eqX}\\
- (\ddot{\phi} + 3H \dot{\phi}) - \epsilon (\dot{X} + 3HX) &=&
W_{,\phi}\,,
\label{eqp}
\end{eqnarray}
where $\gamma_{,X}={\rm d}\gamma/{\rm d} X$, etc. Taking the
time-derivative of Eq.~(\ref{FR}) and using
Eqs.~(\ref{eqX}) and (\ref{eqp}), we obtain
\begin{eqnarray}
\nonumber \dot{H}=-\frac{\kappa^2}{2} \left(\rho+p+\rho_m+\frac43 \rho_r
\right)\,,
\end{eqnarray}
where
\begin{eqnarray}
\label{rhop}
\rho+p=\epsilon \dot{\phi}X+\alpha \dot{H} X^2+2\alpha H
X\dot{X} +\gamma \dot{X}^2 +\dot{\phi}^2 +\alpha_{,X} H X^2 \dot{X}\,.
\end{eqnarray}
In what follows we assume for simplicity that $\alpha$ and $\gamma$ are
constants, i.e., $\alpha_{,X}=\gamma_{,X}=0$.

Following Ref.~\cite{Rubakov} we consider the simplest potential for the
fields,
\begin{eqnarray}
\label{poten}
W(\phi)=\frac12 m^2\phi^2\,,\quad U(X)=-\frac12 M^2 X^2\,,
\end{eqnarray}
which allows for a possibility to realize a phantom phase.

\section{Dynamics of dark energy}
\label{dynamicssec}

One way to analyse the cosmological dynamics in our 
model is to make use of the autonomous equations, 
 the techniques widely
used in the context of dark energy studies~\cite{CLW,CST,GUM}.
This approach is presented in Appendix~B, where we analytically
confirm that our model can lead to the sequence of
radiation, matter and accelerated epochs. 
Also, in Appendix~B we derive the conditions under which 
the de~Sitter solution given below is an attractor. 
Here we first present a simpler analysis based on the slow-roll
approximation. Then we give numerical solutions to eqs.~(\ref{8*}),
(\ref{9*}), (\ref{FR}), (\ref{eqX}), (\ref{eqp}), exhibiting transient 
phantom behaviour, and study their dependence on various parameters
of our model, including the initial values of the fields.

\subsection{Final and initial stages}
\label{sub:prelim}

One immediate point to note is that in the absence of radiation and
matter, the system of equations (\ref{FR}), (\ref{eqX}), (\ref{eqp}) has a
de~Sitter solution, $H={\mathit const}$, for which $\phi$ and $X$ are also
independent of time, provided that
\begin{eqnarray}
\label{epcon}
\frac{\epsilon}{m}>\sqrt{\frac{2\alpha}{3}}\,.
\end{eqnarray}
Indeed, for constant $H,\phi$ and $X$ eqs.~(\ref{FR}), (\ref{eqX}),
(\ref{eqp}) reduce to a simple algebraic system
\begin{eqnarray}
&&H^2=\frac{\kappa^2}{3}\left[-\frac{3\alpha}{2}H^2X^2-\frac{M^2}{2}X^2+
\frac{m^2}{2}\phi^2\right]\,,\nonumber\\
&&3\alpha H^2=M^2\,,\nonumber\\
&&-3\epsilon HX=m^2\phi\,.
\end{eqnarray}
Once the inequality (\ref{epcon}) is satisfied, this system has a solution
\begin{eqnarray}
&&H_{\rm A}=\frac{M}{\sqrt{3\alpha}}\,,\nonumber\\
&&\phi_{\rm A}=\sqrt{\frac{3}{4\pi}} \frac{Mm_{\rm pl}\epsilon
}{\sqrt{\alpha}m^{2}} \frac{1}{\sqrt{3\epsilon ^{2}/m^{2}-2\alpha
}}\,,\nonumber\\
&&X_{\rm A}=-\frac{m_{\rm pl}}{\sqrt{4\pi}}
\frac{1}{\sqrt{3\epsilon^2/m^2-2\alpha}}\,.
\label{phiv}
\end{eqnarray}
We will see in what follows, and elaborate in Appendix~B, 
that in a
range of parameters this solution is an attractor which corresponds to the
de~Sitter phase in asymptotic future (hence the notation). 
In order to use this for dark energy
we require that the mass scale $M$ is of the order of the present Hubble
parameter $H_0$. Then the Newtonian 
effective Planck mass, Eq.~(\ref{newton}), is given by
\begin{eqnarray}
m_{\rm pl, Newton}^2=m_{\rm pl}^2 \left( 1-\frac{\alpha }{3\epsilon^2/
m^2-2\alpha } \right)\,.
\end{eqnarray}
In order that the change of the Planck mass be small, we
impose the condition
\begin{eqnarray}
\label{epco}
\epsilon \gg \sqrt{\alpha}m\,.
\end{eqnarray}
It is worth noting that under this condition, the contribution of the
field $\phi$ in the energy density dominates in the de~Sitter regime,
\begin{equation}
\frac{m^2}{2} \phi_A^2 \gg \frac{M^2}{2} X_A^2 
= \frac{3\alpha H^2}{2}
X_A^2 \; .
\label{page4+}
\end{equation}
Thus, as the system approaches the de~Sitter attractor,
the total energy density in the Universe becomes determined by
the scalar field energy density.

Another point to note is that at early times
(at the radiation-dominated epoch already), when the Hubble parameter is
large enough, the term $(-3\alpha H^2X)$ in Eq.~(\ref{eqX}) drives the
field $X$ to zero, the relevant time being of the order of the Hubble
time. Soon after that the field $\phi$ obeys the usual scalar field
equation in the expanding Universe, so the Hubble friction freezes this field
out. Thus, the initial data for the interesting part of the DE 
evolution are
\begin{eqnarray}
&&X_i=0\,,\nonumber\\
&&\phi_i={\mathit const}\,.
\label{**initial}
\end{eqnarray}
The value of $\phi_i$ is a free parameter of the cosmological evolution in
our model. Since at early times the field $X$ is close to zero,
its effect on the evolution of the field $\phi$ is negligible.
The field $\phi$ slowly rolls {\it down} its potential, and its energy 
density dominates over that of $X$.
Therefore, EOS for DE at early times is normal, $w>-1$, with $w$
being close to $-1$. We refer to this regime as quintessence stage.
As we will see below, in a range of parameters, the system 
eventually crosses the cosmological constant boundary
$w=-1$ and passes through a
transient phantom phase 
before
reaching the de~Sitter asymptotics (\ref{phiv}).

\subsection{Slow roll phantom regime}

The approach to the de~Sitter solution (\ref{phiv}) 
occurs in the slow roll regime.
To see how this happens, we truncate Eqs.~(\ref{eqX}) and
(\ref{eqp}) to
\begin{eqnarray}
\label{phan1}
& &\epsilon \dot{\phi}-3\alpha H^2 X =U_{,X}\,,\\
\label{phan2}
& &-3\epsilon HX =W_{,\phi}\,.
\end{eqnarray}
This truncation is legitimate provided that in addition to the usual
slow-roll conditions $\ddot{\phi} \ll H \dot{\phi}$ and $\ddot{X} \ll H
\dot{X}$, the following conditions are satisfied:
\begin{eqnarray}
\label{con2}
\dot{\phi} \ll \epsilon X \,,\\
\label{con3}
\epsilon \dot{\phi}X \ll V \,,\\
\label{con1}
\dot{X} \ll HX \,.
\end{eqnarray}
[When writing inequalities, we always mean the absolute values of the
quantities.] Note that we do not impose the condition $\epsilon \dot{\phi}
\gg 3\alpha H^2 X$ unlike in Ref.~\cite{Rubakov}, since the term $3\alpha
H^2 X$ is not necessarily negligible relative to the term $U_{, X}$ 
in Eq.~(\ref{phan1}).

{}From Eq.~(\ref{phan1}) we obtain
\begin{equation}
X=-\frac{\epsilon \dot{\phi}}{\xi M^2} \, ,
\label{page6*}
\end{equation}
where
\begin{equation}
\xi \equiv 1-3\alpha \frac{H^2}{M^2}\, .
\label{xi}
\end{equation}
Note that $\xi$ may be considered as a measure of the deviation 
 from the de~Sitter regime (\ref{phiv}).

Substituting Eq.~(\ref{page6*}) 
into Eq.~(\ref{phan2}) we get the following equation
\begin{eqnarray}
\label{Hdsigma}
3H\dot{\phi}=\xi \tilde{W}_{,\phi}\,,
\end{eqnarray}
where
\[
\tilde{W}(\phi) \equiv \frac{m^2M^2}{2\epsilon^2}
\phi^2\,. 
\]
Equation (\ref{Hdsigma}) 
shows that the field $\phi$ rolls {\it up} the potential 
$W(\phi)$ for
$\xi>0$, i.e., for
\begin{eqnarray}
\label{Hcondition}
H<\frac{M}{\sqrt{3\alpha}}\,.
\end{eqnarray}
This is the region in which the phantom equation of state ($w<-1$) is
realized; indeed, Eq.~(\ref{rhop})  gives $\rho+p \approx \epsilon
\dot{\phi}X=\xi X U_{,X} \equiv - \xi M^2X^2$. 
Another way to understand the phantom behaviour is to notice that
when the system approaches the de~Sitter regime, the field $\phi$
dominates the energy density, see Eq.~(\ref{page4+}), so the energy
density increases as the field $\phi$ rolls up.

Let us find out whether the slow roll conditions (\ref{con2}),
(\ref{con3}) and (\ref{con1}) are indeed satisfied. Making use of 
Eq.~(\ref{page6*}) we obtain that the condition (\ref{con2}) is
equivalent to 
\begin{equation}
\epsilon^2 \gg \xi M^2 \, ,
\label{V3*}
\end{equation}
while using Eqs.~(\ref{page6*}) and (\ref{Hdsigma}) we rewrite
the condition (\ref{con3}) as
\begin{equation}
\epsilon^2 \gg \xi \frac{m^2 M^2}{H^2} \; .
\label{V3**}
\end{equation}
The second inequality ensures also the validity of
the relation (\ref{con1}); this can be seen by taking the
time derivative of
Eq.~(\ref{phan2}). The latter two inequalities are automatically
valid at small $\xi$, that is near the de~Sitter solution
(\ref{phiv}).
We conclude that the approach to the de~Sitter solution indeed
occurs
in the slow roll regime, and that the phantom phase is indeed
realised provided that the relation (\ref{Hcondition}) holds.
Our analysis implies also that the de~Sitter solution (\ref{phiv})
is an attractor: the Hubble parameter slowly increases towards
its de~Sitter value, $\xi$ decreases, and
the dynamics gets frozen as $\xi \to 0$.

Since the field $\phi$ dominates  the  energy density at the
phantom slow roll stage, the condition (\ref{Hcondition}) takes 
a simple form 
\begin{equation}
  \phi < \phi_A = \frac{M m_{Pl}}{\sqrt{4 \pi \alpha} m} \, ,
\label{philessphia}
\end{equation}
where we made use of (\ref{epco}). The latter relation 
translates into the range of initial conditions which eventually lead
to the transient phantom behaviour,
\begin{equation}
\phi_i \lesssim \phi_A \,.
\label{a***}
\end{equation}
Indeed, during the radiation- and matter-dominated stages the
field $\phi$ remains almost constant, and at the 
quintessence stage it also does not roll down much.

Recalling again that $\phi$ dominates the energy density,
we rewrite the inequality (\ref{V3**}) as $\varepsilon_s \ll 1$, where
\begin{equation}
\varepsilon _{s} =\frac{2\alpha }{3}
\frac{m^{2}}{\epsilon^{2}}\left(\frac{\phi _{A}}{\phi } 
\right)^{2} \xi\,.
\label{a*}
\end{equation}
The parameter $\varepsilon_s$ may be viewed as the slow roll parameter
for the field $\phi$. Indeed, one observes that
\begin{equation}
   \frac{\dot{\phi}^2}{\xi \tilde{W}} = \varepsilon_s \, ,
\end{equation}
which, together with Eq.~(\ref{Hdsigma}), justifies this
interpretation.

{} From Eqs.~(\ref{page3+}), (\ref{rhop}),  (\ref{page6*}) and  
(\ref{Hdsigma}) it follows that during the slow roll phantom stage,
the EOM parameter of DE is given by
\begin{equation}
w = -1 - \varepsilon_s \, .
\label{a**}
\end{equation}
Hence, the appreciable deviation from $w=-1$ occurs when
$\varepsilon_s$ is not much smaller than unity, i.e., when
$\phi_i$ is appreciably smaller than $\phi_A$.

In the next section we confirm these expectations by numerical
analysis, and also show explicitly that in a range of parameters,
the cosmological evolution proceeds from radiation-dominated to
matter-dominated epoch,
and then to the slow roll phantom stage, before finally ending up in
the de~Sitter regime (\ref{phiv}).

\subsection{Numerical solutions}

In our numerical analysis we choose initial
conditions $\dot X = \dot \phi = X = 0$ with
nonzero values of $\phi$, $\rho_m$ and
$\rho_r$. This choice corresponds to the initial data (\ref{**initial}).
We have also tried many other initial conditions and found that the
results are not sensitive to the initial values of $X$, $\dot{X}$ and
$\dot{\phi}$, in accord with the discussion in the end of
Sec.~\ref{sub:prelim}.

\begin{figure}[H]
\begin{center}
\includegraphics[width=3.8in,height=3.8in]{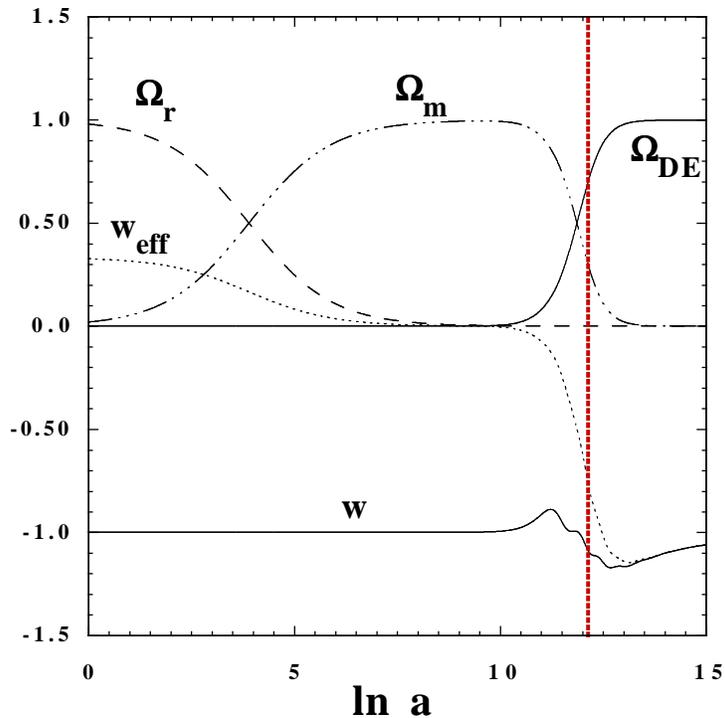}
\caption{\label{evo1}
Cosmological evolution  for the model parameters
$\alpha=1$, $\gamma=1/2$, $\epsilon/m=3$ and $M/m=1$. We choose initial
conditions $X=\dot{X}=\dot{\phi}=0$,  $\phi_i = 0.5 \phi_A$, 
and $\Omega_{r, i}=0.99$, $\Omega_{m,i} = 0.01$. Shown is 
the evolution of
$\Omega_{\rm DE}$, $\Omega_m$, $\Omega_r$, $w$ 
and $w_{\rm eff}$ as functions of  $N\equiv \ln a$.
Note that the present epoch corresponds to $\Omega_m = 0.3$
and $\Omega_{\rm DE}=0.7$, which is denoted by a vertical line.
After the cosmological constant boundary crossing, 
the DE EOS parameter $w$ reaches a minimum 
$w_{\rm min}=-1.19$ and then increases towards 
the de~Sitter value $w=-1$ from the phantom side.}
\end{center}
\end{figure}

In Fig.~\ref{evo1} we plot the cosmological evolution for the model
parameters $\alpha=1$, $\gamma=1/2$, $\epsilon /m=3$,
$M/m=1$ and the initial value $\phi_i = 0.5 \phi_A$. 
We find it convenient to present the plots in terms of the variable
\begin{equation}
N = \ln a \, .
\end{equation}
Figure~\ref{evo1} clearly shows that the sequence of radiation, 
matter and de~Sitter
epochs can be achieved together with the $w=-1$ crossing. The DE EOS
parameter $w$ is nearly a constant, $w \simeq -1$, during the radiation
and matter epochs because the fields are almost frozen. At the 
transition era from matter domination to DE domination,
$w$ begins to grow because the kinetic energies of the fields become
important; this is the quintessence phase.
 However, the system soon enters the phantom phase during which
the field $\phi$ rolls up the potential. Hence the equation of state $w$
crosses the cosmological constant boundary $w=-1$ and reaches a minimum
value $w_{\rm min}<-1$. The solution finally approaches the de~Sitter
regime (\ref{phiv}) from the phantom side. 
Of some interest is also the EOS parameter of the entire system,
\begin{equation}
w_{\rm eff}=\frac{p_{\rm tot}}{\rho_{\rm tot}}\,.
\end{equation}
It is seen from Fig.~\ref{evo1}, this parameter also becomes smaller
than $-1$ soon after $w=-1$ crossing. 

The pattern shown in Fig.~\ref{evo1} is generic in our model, provided
that its parameters and initial data obey $m/\epsilon \ll 1$, $M/\epsilon
\ll 1$ and $\phi_i < \phi_A$ (in fact, the inequalities here need not be
strong). The strengths of the effects depend, of course, on the values of
these parameters. In particular, the minimum value $w_{\rm min}$ is
related to the slow-roll parameter $\varepsilon_s$, in accord with
Eq.~(\ref{a**}). 
If the field $\phi$ evolves very slowly, one has $\varepsilon_s \ll 1$, 
so $w_{\rm min}$ is close to $-1$. On the contrary, the appreciable 
deviation from $w=-1$
occurs if $\varepsilon_s$ is not very much smaller than unity.

\begin{figure}[H]
\begin{center}
\includegraphics[width=3.8in,height=3.8in]{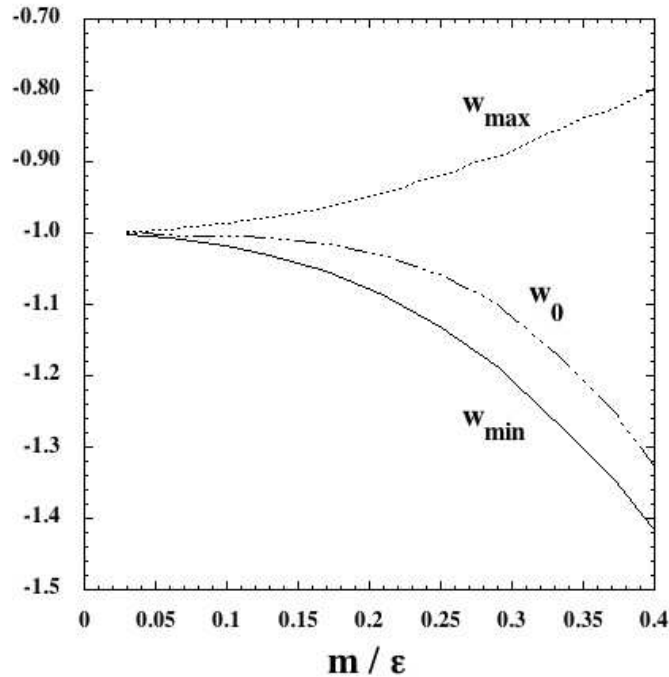}
\caption{\label{A}
The minimum value of the EOS parameter $w$ of DE, its value $w_{0}$ at the
present epoch ($\Omega _{m}=0.3$, $\Omega _{\rm DE}=0.7$) and its maximum
value as functions of $m/\epsilon$ for $M/\epsilon= 1/30$ and the initial
value $\phi_i= 0.5 \phi_A$.}
\end{center}
\end{figure}

Since the field $\phi$ is practically frozen during the radiation- and
matter-dominated epochs, and evolves rather slowly later on,
the dependence of the cosmological evolution on the parameters of the
model and on the initial value $\phi_i$ can be understood, at qualitative
level, by inspecting Eq.~(\ref{a*}). For the qualitative discussion
of the evolution well before the asymptotic de~Sitter regime sets in,
the parameter $\xi$ in (\ref{a*}) may be set equal to 1, while the
value of $\phi$ may be set equal to its initial value $\phi_i$.
Then Eq.~(\ref{a*}) implies that with other parameters and $\phi_i$ fixed,
for smaller $m/\epsilon$ one gets
smaller absolute value of $\varepsilon_s$ at its minimum, leading to the
value of $w_{\rm min}$ closer to $-1$. This is shown in Fig.~\ref{A}.
{}From Fig.~\ref{evo1} it is clear, however, that the minimum of $w$
occurs after the present epoch ($\Omega_{\rm DE} \simeq 0.7$ and 
$\Omega_{m} \simeq 0.3$).
Again, this is a rather generic feature of our model. Therefore,
instead of $w_{\rm min}$, more interesting quantities are the present value
$w_0$ of the DE EOS parameter and also its maximum value before the
cosmological constant boundary crossing. These quantities are also given
in Fig.~\ref{A}. 
Overall, the behaviour shown in Fig.~\ref{evo1} is more
pronounced at larger $m/\epsilon$,
once other parameters of solutions are kept fixed.

\begin{figure}[H]
\begin{center}
\includegraphics[width=3.8in,height=3.8in]{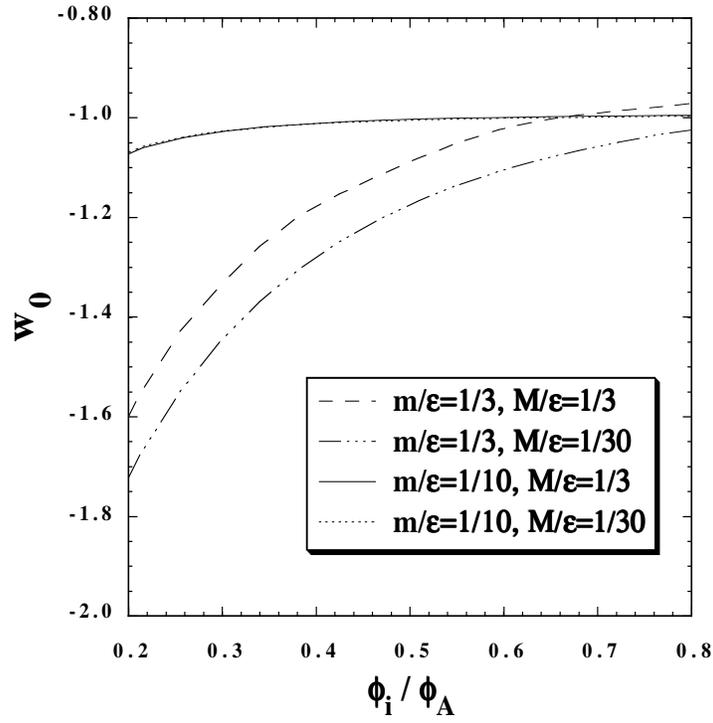}
\caption{\label{D}
The dependence of the present value $w_0$ (i.e., $w$ at $\Omega_m = 0.3$,
$\Omega_{\rm DE}=0.7$) on the initial value of $\phi$, 
for different sets of the model parameters.}
\end{center}
\end{figure}

\begin{figure}[H]
\begin{center}
\includegraphics[width=3.8in,height=3.8in]{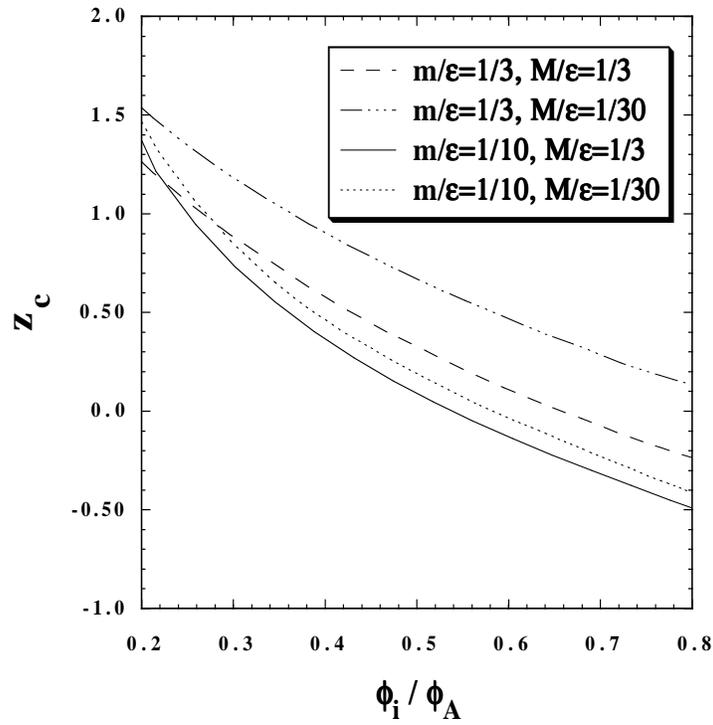}
\caption{\label{E}
The dependence of the redshift (from the present epoch, $\Omega_m = 0.3$,
$\Omega_{\rm DE}=0.7$) of the cosmological constant boundary crossing, $w=-1$,
on the initial value of $\phi$.}
\end{center}
\end{figure}

According to Eq.~(\ref{a*}),
the initial  value of the field $\phi$
is also important to determine the amplitude of $\varepsilon_s$ and hence
$w$: a smaller
$\phi_i$ results in a stronger deviation of $w_{\rm min}$
from $-1$. The present value $w_0$ also becomes more negative (stronger
deviating from $-1$), while the cosmological constant boundary crossing
occurs earlier.  From the numerical analysis we find that
at relatively large values of $m/\epsilon$ the increase of
$M$ has the opposite effect, while at smaller $m/\epsilon$ the effects due
to the variation of $M$ are small. These properties are illustrated in
Figs.~\ref{D} and \ref{E}.

\begin{figure}[H]
\begin{center}
\includegraphics[width=3.6in,height=3.6in]{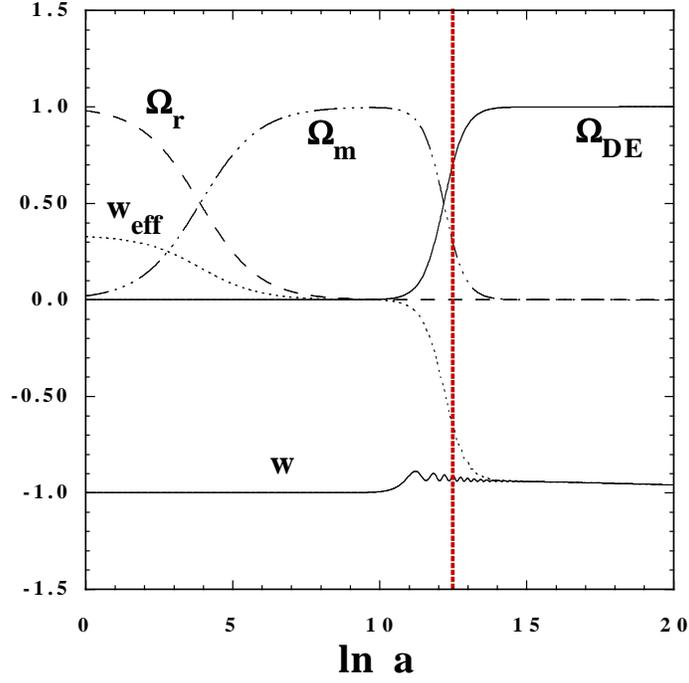}
\caption{\label{evo3}
Cosmological evolution for the model parameters $\alpha=1$, $\gamma=1/2$,
$\epsilon/m=3$ and $M/m=0.1$. 
We choose initial conditions $X=\dot{X}=\dot{\phi}=0$,  $\phi_i = 3\phi_A$, 
and $\Omega_{r, i}=0.99$, $\Omega_{m,i} = 0.01$.
The present epoch corresponds to $\Omega_{DE}= 0.7$
and $\Omega_m = 0.3$, which is denoted by a vertical line.
In this case the cosmological constant boundary crossing
is not realized because $\phi$ remains always 
larger than $\phi_A$.}
\end{center}
\end{figure}

As discussed above, the phantom phase occurs only if the initial value
of $\phi$ obeys (\ref{a***}), otherwise the de~Sitter attractor is
approached from the quintessence side, $w>-1$. 
This is illustrated in Fig.~\ref{evo3}.


At the end of this section we discuss the variation of the
effective gravitational constant. According to
Eq.~(\ref{newton}), the effective Newton's constant that determines the
interaction deep inside the horizon scale is given by
\begin{eqnarray}
G_{*}=G(1-4\pi G \alpha X^2)^{-1}\,.
\label{Eqn/Pg18/1:LV16}
\end{eqnarray}
Its variation in time is conveniently expressed in terms of the following
quantity
\begin{eqnarray}
\frac{{\rm d} \ln G_{*}}{{\rm d} \ln a} 
\equiv \frac{\dot{G}_*}{G_* H} = \frac{8\pi \alpha X
\dot{X}} {H (m_{\rm pl}^2 - 4\pi \alpha X^2)} \, .
\end{eqnarray}
The typical experimental and observational constraints on the variation of
$G_*$ in the present Universe are given by $|\dot{G_*}/G_*| \lesssim
10^{-12}\,{\rm yr}^{-1}$ \cite{Uzan}, which translates into the condition
\begin{eqnarray}
\label{congra}
\left| \dot{G_*}/G_* \right| \lesssim 10^{-2} H_{0}\, .
\end{eqnarray}
\begin{figure}[H]
\begin{center}
\includegraphics[width=3.5in,height=3.5in]{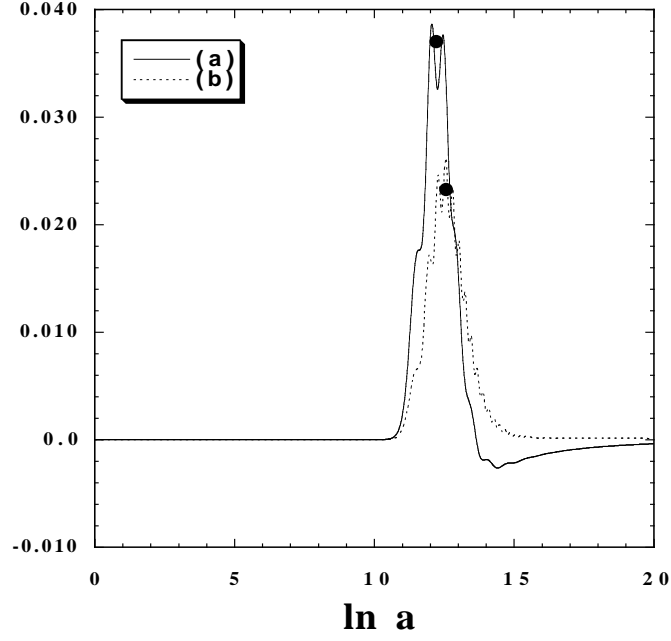}
\caption{\label{gravitation}
The evolution of the variation of the effective gravitational constant
${\rm d} \ln G_{*}/{\rm d} \ln a$ as function of $N=\ln a$ and redshift $z$. 
The cases (a)  and (b) correspond to the model parameters
and initial conditions given of Figs.~\ref{evo1} and \ref{evo3},
respectively. The black points represent the values at the present epoch
($\Omega_m \simeq 0.3$). }
\end{center}
\end{figure}
\begin{figure}[H]
\begin{center}
\includegraphics[width=3.8in,height=3.8in]{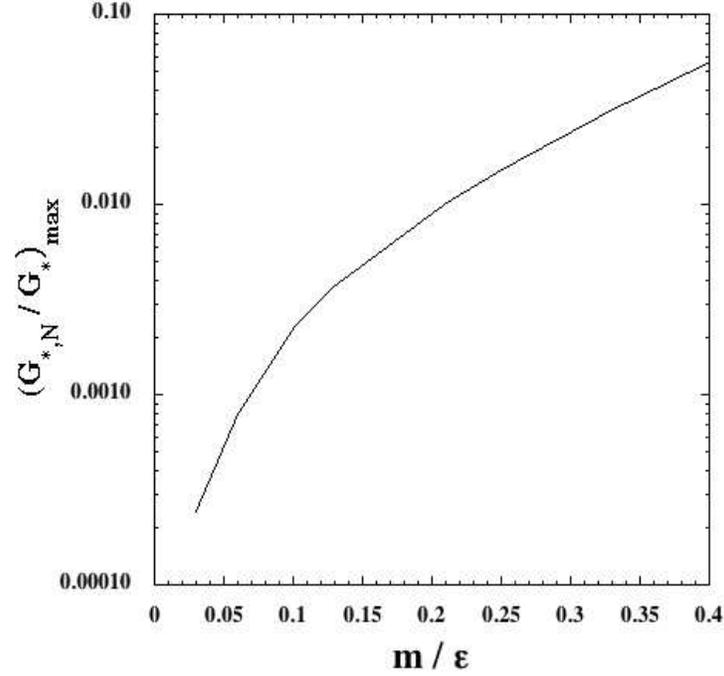}
\caption{\label{F}
The maximum value of ${\rm d} \ln G_{*}/{\rm d} \ln a$ 
as a function of $m/\epsilon$. Other
parameters and initial conditions are the same as in Fig.~\ref{A}. }
\end{center}
\end{figure}

In Fig.~\ref{gravitation} we plot the evolution of the quantity
${\rm d} \ln G_{*}/{\rm d} \ln a$ 
for the model parameters and initial conditions given in
Figs.~\ref{evo1} and \ref{evo3}. At the present epoch ($\Omega_m \simeq
0.3$) we obtain the values $\dot{G_*}/G_*=3.5 \times 10^{-2}H_0^{-1}$ and
$2.4 \times 10^{-2}H_0^{-1}$ for these two cases, respectively.
Comparing Fig.~\ref{gravitation} with 
Figs.~\ref{evo1} and \ref{evo3} one observes that the variation of the
gravitational constant is correlated in time with the deviation of $w$
from $-1$. This is clear from Eq.~(\ref{Eqn/Pg18/1:LV16}) too: the
gravitational constant varies when the field $X$ changes in time, while
the latter occurs during the transition from the matter dominated stage to
the final de~Sitter attractor. It is precisely at this
transition stage that $w$ substantially deviates from $-1$.

Figure~\ref{F} shows the maximum value of ${\rm d} \ln G_*/{\rm d} \ln a$ 
as a function of $m/\epsilon$. Again, the variation of the effective Newton's
constant is more pronounced at larger $m/\epsilon$. This means that it is
correlated with the amplitude of the deviation of $w$ from $-1$. The
dependence of ${\rm d} \ln G_* /{\rm d} \ln a$ on $M$ and on the
initial value of $\phi$ is rather weak.

It is worth pointing out that as long as the deviation of the EOS from
$w=-1$ is not so significant, the models satisfy the constraint
(\ref{congra}), and also that our model suggests that the variation of $G_*$
is close to the present upper bound on $\dot{G_*}/G_*$. 

\section{Momentum scales of instabilities}
\label{instabilitysec}

In this section the momentum scales of instabilities are present in
our model. We first study dispersion relations in  the Minkowski space-time
and then proceed to those in the FRW space-time.
We wish to clarify the  conditions under which a tachyon or a superluminal mode
appears by considering dispersion relations. 
We also evaluate the energy of the modes to find out a ghost state. 

\subsection{Minkowski spectrum}

Let us consider the perturbations for the fields,
\begin{eqnarray}
B_0=X+b_0\,, \quad B_i=b_i\,, \quad \Phi=\phi+\varphi\,.
\end{eqnarray}
The quadratic Lagrangian  for perturbations, following from the general
expression (\ref{lag}), is
\begin{eqnarray}
\label{Lag2}
{\cal L}_{b_0, b_i, \varphi}&=& \frac{\gamma}{2} \partial_\mu
b_0 \partial^{\mu}b_0 +\frac{\alpha}{2} \partial_\mu b_i \partial^{\mu}b_i
+\frac12 \partial_\mu \varphi \partial^{\mu}\varphi +\epsilon \partial_0
\varphi b_0-\epsilon \partial_i \varphi b_i\nonumber\\
&& -\frac12 m_0^2 b_0^2-\frac12 m_1^2 b_i^2 -\frac12
m_{\varphi}^2\varphi^2\,,
\end{eqnarray}
where
\begin{eqnarray}
m_0^2=U_{,XX}\,, \quad m_1^2=-\frac{U_{,X}}{X}\,, \quad
m_{\varphi}^2=W_{,\phi \phi}\,.
\end{eqnarray}
For our model (\ref{poten}) one has $-m_0^2=m_1^2=M^2$ and
$m_{\varphi}^2=m^2$. In what follows we concentrate on this case, and
assume the following relations, see (\ref{epco}) and
(\ref{V3*}),
\begin{eqnarray}
& &\epsilon \gg \sqrt{\alpha }m\,,\nonumber\\
& &\epsilon \gg M\,.
\label{Eq/Pg21/1:draft-apr6}
\end{eqnarray}

Varying the Lagrangian (\ref{Lag2}) with respect to  $b_i$, $b_0$ and
$\varphi$, we obtain the equations for the field perturbations. In order
to find the spectrum of the system we write the solutions in the form
$b_0=\tilde{b}_0 e^{i {\bf p}_{\mu} {\bf x}^{\mu}} =\tilde{b}_0 e^{i
(\omega t- {\bf p}\cdot {\bf r})}$, $b_i=\tilde{b}_i e^{i {\bf p}_{\mu}
{\bf x}^{\mu}}$ and $\varphi=\tilde{\varphi} e^{i {\bf p}_{\mu} {\bf
x}^{\mu}}$. The transverse mode of the vector field $B_i$ has the
dispersion relation
\begin{eqnarray}
\omega_0^2=p^2+\frac{M^2}{\alpha}\,.
\end{eqnarray}
The three scalar modes $\tilde{b}_i=(p_i/p)\tilde{b}_L$, $\tilde{b}_0$ and
$\tilde{\varphi}$ satisfy the following equations
\begin{eqnarray}
\label{spe1}
& & \left( \omega^2-p^2-\frac{M^2}{\alpha} \right)
\tilde{b}_L+i\frac{\epsilon}{\alpha}p\tilde{\varphi}=0\,,\\
\label{spe2}
& & \left( \omega^2-p^2+\frac{M^2}{\gamma} \right)
\tilde{b}_0 -i\frac{\epsilon}{\gamma}\omega \tilde{\varphi}=0\,,\\
\label{spe3}
& & \left( \omega^2-p^2-m^2 \right) \tilde{\varphi}
-i\epsilon \omega \tilde{b}_0-i\epsilon p \tilde{b}_L=0\,.
\end{eqnarray}
Expressing $\tilde{b}_L$ and $\tilde{b}_0$ in terms of $\tilde{\varphi}$
from Eqs.\,(\ref{spe1}), (\ref{spe2}) and plugging them into
Eq.~(\ref{spe3}), we find that the eigenfrequencies corresponding to three
mixed states satisfy
\begin{eqnarray}
\label{eqx}
(z-m^2)\left(z +\frac{M^2}{\gamma} \right) \left(z
-\frac{M^2}{\alpha} \right) -\epsilon^2 z \left( \frac{z}{\gamma}+
\frac{p^2}{\gamma}+\frac{p^2}{\alpha} -\frac{M^2}{\gamma \alpha}
\right)=0\,,
\end{eqnarray}
where
\begin{eqnarray}
z \equiv \omega^2 -p^2\,.
\end{eqnarray}
The spectrum in the case $m=0$ was studied in Ref.~\cite{Rubakov}. Our
purpose here is to extend the analysis to the case of non-zero $m$.
Denoting the solutions of Eq.~(\ref{eqx}) as $z_1$, $z_2$ and $z_3$, we
obtain the relation
\begin{eqnarray}
\label{xrelation}
z_1 z_2 z_3=-\frac{m^2M^4}{\alpha \gamma}\,.
\end{eqnarray}
Once the conditions (\ref{Eq/Pg21/1:draft-apr6}) are satisfied, then one
can show that if the relation
\begin{eqnarray}
\label{ineq}
z_{1}<z_{2}<z_{3}
\end{eqnarray}
holds at some momentum, then the inequality (\ref{ineq}) is satisfied for
all momenta.

In the limits $p \to \infty$ and $p \to 0$, we obtain the following
dispersion relations, respectively.
\begin{itemize}
\item (A) UV limit ($p \to \infty$)
\begin{eqnarray}
\label{Spectrum1}
\omega _{1}&=&p-\frac{\epsilon }{2}\sqrt{\frac{1}{\gamma
}+\frac{1}{\alpha }}+\frac{\epsilon ^{2}}{8p}\left(\frac{2m^{2}}{\epsilon
^{2}}+\frac{1}{\gamma }-\frac{1}{\alpha } \right)+\mathcal{ O}(M^2/p)\,,\\
\nonumber
\omega _{2}&=&p+\frac{m^{2}M^{4}}{2p^{3}\epsilon ^{2}(\alpha +\gamma
)}+\mathcal{ O}(1/p^{5})\,,\\
\nonumber
\omega _{3}&=&p+\frac{\epsilon }{2}\sqrt{\frac{1}{\gamma }+\frac{1}{\alpha
}}+\frac{\epsilon ^{2}}{8p}\left(\frac{2m^{2}}{\epsilon
^{2}}+\frac{1}{\gamma }-\frac{1}{\alpha } \right)+\mathcal{ O}(M^2/p)\,.
\end{eqnarray}
We see that $\omega _{1}<\omega _{2}<\omega _{3}$ and $z_{1}<0$,
$z_{2,3}>0$. In all three cases the group velocities $\partial \omega
_{i}/\partial p$ are less than 1, so neither mode is superluminal at high
three-momenta, provided that $\alpha >\gamma $. The two-derivative terms
in the Lagrangian (\ref{Lag2}) dominate in the UV limit, so there are
neither ghosts nor tachyons in this limit.
\item (B) IR limit ($p \to 0$)
\begin{eqnarray}
\omega _{1}&=&-\frac{m^{2}M^{2}}{\epsilon ^{2}}
\left[1+\mathcal{O}(m^2/\epsilon^2, M^2/\epsilon^2) \right]\,,
\label{Spectrum01}\\
\omega _{2}&=&\frac{M^{2}}{\alpha}\,, \nonumber
\\
\omega _{3}&=&\frac{\epsilon ^{2}}{\gamma } +m^{2}+\mathcal{
O}(M^{2})\,. \nonumber
\end{eqnarray}
We see again that $\omega _{1}<\omega _{2}<\omega _{3}$ and $z_{1}<0$,
$z_{2,3}>0$. This means that using the property (\ref{ineq}) we can
identify the modes: the first one has the behaviour (\ref{Spectrum1}) and
(\ref{Spectrum01}), and so on.
\end{itemize}

It follows from Eq.~(\ref{xrelation}) that $z_{i}$ never vanish. In fact
the coefficients in Eq.~(\ref{eqx}) are regular at all momenta, so $z_{i}$
are regular as well. Therefore, $z_{i}$ never change signs and hence
$z_{1}<0,\ z_{2,3}>0$ for all momenta. This means, in particular, that the
second and third modes never become tachyonic.

Let us discuss the dangerous mode with the dispersion relation $\omega
=\omega _{1}(p)$ in some detail.
The expression for the fields in each mode is
\begin{eqnarray}
\tilde{b}_{L,i}&=&-i\epsilon p\left(\gamma z_{i}
+M^{2} \right)\cdot C_{i}\,,
\nonumber\\
\tilde{b}_{0,i}&=&i\epsilon \omega
(\alpha z_{i}-M^{2})\cdot C_{i}\,,
\nonumber\\
\tilde{\varphi}_{i}&=&(\gamma z_{i}+M^{2})
(\alpha z_{i}-M^{2})\cdot C_{i}\,,
\label{Eq/Pg24/1:draft-apr6}
\end{eqnarray}
where $C_{i}$ are the normalization factors. Setting $\omega^2=0$ in
Eq.~(\ref{eqx}), we obtain three critical momenta
\begin{eqnarray}
& &p_{1, 2}^2=\frac12 \left[ \frac{\epsilon^2-M^2} {\alpha}-m^2 \pm
\sqrt{\left(\frac{\epsilon^2-M^2}{\alpha}-m^2\right)^2
-\frac{4m^2M^2}{\alpha}} \right]\,, \\
& & p_3^2=\frac{M^2}{\gamma}\,.
\end{eqnarray}
Under the conditions (\ref{Eq/Pg21/1:draft-apr6}), the critical momenta
$p_{1,2}^2$ are approximately given by
\begin{eqnarray}
p_1^2 \simeq \frac{\epsilon^2-M^2}{\alpha}-m^2\,, \quad p_2^2 \simeq
\frac{m^2M^2}{\epsilon^2}\,,
\end{eqnarray}
so that $p_1^2>p_3^2>p_2^2>0$. The tachyonic mode ($\omega_1^2<0$) is
present for $0<p^2<p_2^2$ and $p_3^2<p^2<p_1^2$.

In order to find whether there are ghosts we calculate the energy of the
modes (\ref{Eq/Pg24/1:draft-apr6}),
\begin{eqnarray}
E_{i}(p)=2\omega ^{2}|C_{i}|^{2}\left[\alpha \epsilon ^{2}p^{2}(\gamma
z_{i}+M^{2})^{2}+\epsilon ^{2}(\gamma p^{2}-M^{2})(\alpha z_{i}-M^{2})^{2}
+(\gamma z_{i}+M^{2})^{2}(\alpha z_{i}-M^{2})^{2} \right]\,.
\end{eqnarray}
For the modes with $\omega _{2,3}$ we have $E_{2,3}(p)>0$. For the mode
with $\omega_1$ the energy is equal to zero at $p=p_{3}=M/\sqrt{\gamma}$.
 While $\omega_1^2>0$ for $p^{2}<p_{3}^{2}\equiv M^{2}/\gamma$, the energy
 $E_{1}(p)$ changes its sign at this momentum. Thus the mode with
$\omega_1$ is a ghost for $p^{2}<M^{2}/\gamma$.

We summarize the properties of the dangerous mode as follows:
\begin{itemize}
\item (i) $p^2>(\epsilon^2-M^2)/\alpha-m^2$:~healthy
\item (ii) $M^2/\gamma<p^2<(\epsilon^2-M^2)/\alpha-m^2$:~tachyon
\item (iii) $m^2M^2/\epsilon^2<p^2<M^2/\gamma$:~ghost, but not tachyon
\item (iv) $0<p^2<m^2M^2/\epsilon^2$:~tachyon.
\end{itemize}
Unlike the case $m=0$ \cite{Rubakov} the tachyon is present in the deep IR
region (iv).

To end up the discussion of the modes in Minkowski space-time, we give the
expressions for the minimum values of $\omega ^{2}$ in the tachyonic
regions,
\begin{itemize}
\item (ii):
\begin{equation}
\omega ^{2}_{\mathrm{min}}=-\displaystyle\frac{\gamma \epsilon
^{2}}{4\alpha (\alpha +\gamma )}\ \ \ \mbox{at}\ \ \ p^{2}=\frac{\epsilon
^{2}}{4\alpha }\frac{\gamma +2\alpha }{\gamma +\alpha}\,.
\label{Eq/Pg25/1:draft-apr6}
\end{equation}
\item (iv):
\[
\omega ^{2}_{\mathrm{min}}=-\displaystyle\frac{m^{2}M^{2}}{\epsilon
^{2}}\ \ \ \mbox{at}\ \ \ p^{2}=0\,.
\]
\end{itemize}
Note that $|\omega _{\mathrm{min}}^{2}|$ is relatively large in the region
(ii), so this region is the most problematic.

\subsection{Evolution of perturbations in  cosmological background}

Finally we discuss the evolution of field perturbations in the FRW
background (\ref{FRW}). In the cosmological context the physical momentum
$p$ is related to the comoving momentum $k$ as $p=k/a$. Once the
parameters of the model and initial data are such that the cosmological
boundary crossing occurs, the present epoch ($\Omega_{\rm DE} \simeq 0.7$)
typically corresponds to the phantom region. {}From Fig.\,1  one can see that
the Hubble parameter
does not change much during the transition from the
phantom epoch to the final de~Sitter era. Hence the present value of the
Hubble parameter ($H_0$) is of the same order as the value
$H=M/\sqrt{3\alpha}$ in the de~Sitter asymptotics. This means that the value
$p_{3}=M/\sqrt{\gamma}$ is of the same order as $H_0$ provided that
$\gamma$ and $\alpha$ are of order unity.

The tachyon appears when the momentum  $p=k/a$ of the dangerous mode
becomes smaller than $\sqrt{(\epsilon^2-M^2)/\alpha-m^2}$ and temporally
disappears when the mode crosses the value $M/\sqrt{\gamma}$. Hence this
instability is present for the modes which are inside the Hubble radius
and satisfy $M/\sqrt{\gamma}<p<\sqrt{(\epsilon^2-M^2)/\alpha-m^2}$, but it
is absent for the modes deep inside the Hubble radius, satisfying
$p>\sqrt{(\epsilon^2-M^2)/\alpha-m^2}$. After the Hubble radius crossing
($k=aH$), the tachyonic instability disappears in the momentum region
$m^2M^2/\epsilon^2<p^2<M^2/\gamma$, but the tachyon appears again for
$p^2<m^2M^2/\epsilon^2$. Note that the ghost existing at
$m^2M^2/\epsilon^2<p^2<M^2/\gamma$ is a not a problem
because of its low energy~\cite{Cline,Dubovsky:2004sg}.

In what follows we discuss the evolution of field perturbations in the two
tachyonic regimes. Before doing that it is instructive to study the
high-momentum regime that sets the initial data for the tachyonic
evolution.

\subsubsection{$p^2 \gg(\epsilon^2-M^2)/\alpha-m^2$}

We denote the overall amplitude of the dangerous mode as $\varphi $. Since
the modes are deep inside the Hubble radius ($k/a \gg H$) in the regime we
discuss here, the field perturbation $\chi $ approximately satisfies
\begin{eqnarray}
\frac{{\rm d}^2}{{\rm d}\eta^2}\chi+k^2\chi \simeq 0\,,
\end{eqnarray}
where $\chi=a\varphi$ and $\eta$ is conformal time defined by
$\eta=\int a^{-1}{\rm d}t$.
Taking the asymptotic Minkowski
vacuum state, $\chi=e^{-ik\eta}/\sqrt{2k}$, the squared amplitude of the
field perturbation $\varphi$ is given by \cite{BTW}
\begin{eqnarray}
{\cal P}_{\varphi}=\frac{4\pi k^3}{(2\pi)^3}
|\varphi|^2=\left(\frac{k}{2\pi a}\right)^2\,.
\end{eqnarray}
Since the maximum momentum at which the tachyon appears is $k/a \simeq
\epsilon/\sqrt{\alpha}$, one has the following estimate for the amplitude
of the field perturbation at the beginning of the tachyonic instability,
\begin{eqnarray}
\label{varphii}
\varphi_i \simeq \frac{\epsilon}{2\pi \sqrt{\alpha}}\,.
\end{eqnarray}
As usual, this amplitude characterizes the contribution of a logarithmic
interval of momenta into $\langle \varphi ^{2}(\mathbf{x})\rangle $.

\subsubsection{$M^2/\gamma<p^2<(\epsilon^2-M^2)/\alpha-m^2$}

This interval of momenta is dangerous, as the perturbations undergo the
tachyonic amplification. Since the modes are still inside the Hubble
radius, one can neglect the gravitational effects on the ``frequency''
$\omega$ when  estimating the growth of field perturbations. By the time
the tachyonic amplification ends up, the amplitude of field perturbations
is estimated as
\begin{eqnarray}
\label{growth}
\varphi \simeq \varphi_i \exp \left( \int_{t_i}^{t_f}
|\omega_1| {\rm d} t \right)=\varphi_i \exp \left( \int_{p_1}^{p_3}
\frac{|\omega_1|}{H} \frac{{\rm d}p}{p}\right)\,.
\end{eqnarray}
Recall that $p_1 \simeq \epsilon/\sqrt{\alpha}$ and $p_3=M/\sqrt{\gamma}$.
The largest value of $|\omega_1^2|$ is approximately given by
(\ref{Eq/Pg25/1:draft-apr6}). Substituting this value into
Eq.~(\ref{growth}) and recalling that the background changes slowly
($H \simeq {\mathit const}$),
one finds that the amplitude of the field perturbation
after exit from the tachyonic regime is of order
\begin{eqnarray}
\varphi \simeq \frac{\epsilon}{2\pi \sqrt{\alpha}} \exp \left[ \frac12
\sqrt{\frac{\gamma} {\alpha (\gamma+\alpha)}} \frac{\epsilon}{H} \log
\left( \sqrt{\frac{\gamma}{\alpha}} \frac{\epsilon}{M} \right) \right]\,,
\end{eqnarray}
where we used Eq.~(\ref{varphii}).

Recall now that $H$ is of the same order as $M/\sqrt{3\alpha}$  during the
phantom phase. Hence the large ratio $\epsilon/M$ leads to a strong
amplification of field perturbations. From (\ref{phiv}), the
homogeneous field $\phi$ at the phantom and  de~Sitter phase is estimated as $\phi \simeq
\sqrt{\frac{3}{4\pi}}\frac{Hm_{\rm pl}}{m}$. The requirement that the
perturbation $\varphi$ is smaller than the background field $\phi$ leads
to the constraint
\begin{eqnarray}
\label{notexceed}
\exp \left[ \frac12 \sqrt{\frac{\gamma} {\alpha
(\gamma+\alpha)}} \frac{\epsilon}{H} \log \left(
\sqrt{\frac{\gamma}{\alpha}} \frac{\epsilon}{M} \right) \right]<
\sqrt{3\alpha \pi} \frac{Hm_{\rm pl}}{\epsilon m}\,.
\end{eqnarray}
As an example, in the case $m=H_0=10^{-42}$ GeV, $M=\sqrt{3\alpha }H_{0}$,
$\alpha=1$ and $\gamma=1/2$, we obtain the constraint $\epsilon/M \lesssim
70$. As long as $\alpha$ and $\gamma$ are of order one, the ratio
$\epsilon/M$ should not be too much larger than unity.

\subsubsection{$0<p^2<m^2M^2/\epsilon^2$}

After the Hubble radius crossing, the effect of the cosmic expansion can
no longer be neglected when estimating the ``frequencies'' of the field
perturbations. Since there are no tachyonic instabilities for
$m^2M^2/\epsilon^2<p^2<M^2/\gamma$, we  consider the evolution of
perturbations in the region $0<p^2<m^2M^2/\epsilon^2$. In
Ref.~\cite{Rubakov} the equations for the field perturbations were derived
in the slow-rolling background under the condition $p^2 \ll M^2, m^2$.
The equation for the
perturbation $\chi=a\varphi$ is approximately given by
\begin{eqnarray}
\label{chieq}
\frac{{\rm d}^2}{{\rm d}\eta^2}\chi
+\left( k^2 -\frac{1}{a}\frac{{\rm d}^2 a}
{{\rm d} \eta^2}-a^2
\frac{m^2M^2}{\epsilon^2} \right) \chi=0\,.
\end{eqnarray}
Here we neglected the contribution of metric perturbations 
on the r.h.s. of this equation. Note that metric perturbations works 
as a back reaction effect after the field perturbation is sufficiently 
amplified. The growth rate of the perturbation $\chi$ is mainly 
determined by the terms in the parenthesis of Eq.~(\ref{chieq})
rather than the backreaction of metric perturbations.

The last term corresponds to the tachyonic mass term, which already
appeared in Minkowski spacetime [see Eq.~(\ref{Spectrum01})]. Since the
term $({\rm d}^2a/{\rm d}\eta^2)/a$ is of order $a^2H^2$,
one can estimate the ratio of the
tachyonic mass relative to this gravitational term:
\begin{eqnarray}
\delta \equiv \frac{a^2m^2M^2/\epsilon^2}
{({\rm d}^2a/{\rm d}\eta^2)/a} \simeq
\frac{m^2M^2}{\epsilon^2 H^2}\,.
\end{eqnarray}
If we use the de~Sitter value $H=M/\sqrt{3\alpha}$, this ratio is
estimated as $\delta=3\alpha m^2/\epsilon^2 \ll 1$. Hence the
gravitational term $({\rm d}^2a/{\rm d}\eta^2)/a$
dominates over the tachyonic mass.

In the de~Sitter background with $a=-1/(H\eta)$ the approximate solutions
to Eq.~(\ref{chieq}) can be obtained by setting
$\chi=-(C/H)\eta^{-1+\tilde{\delta}}$. One finds that
$\tilde{\delta}=-m^2M^2/ (3H^2\epsilon^2)$ for the growing solution,
thereby giving
\begin{eqnarray}
\varphi=C\eta^{-\frac{m^2M^2}{3H^2\epsilon^2}} \propto a^{\delta/3}\,.
\end{eqnarray}
In Ref.~\cite{Rubakov} it was shown that the physical temporal component
of the vector field perturbations evolves as $b_0/a \propto a^{\delta/3}$,
whereas the physical spatial component of the vector field decreases as
$B_i/a \propto a^{-1+\delta/3}$. The growth rate of $\varphi$ and $b_0/a$
is small due to the condition $\delta \ll 1$. So, the second tachyonic
instability is harmless for the past and present cosmological evolution.
However, we notice that since the de~Sitter solution is a late-time
attractor, the perturbations $\varphi$ and $b_0/a$ become larger than the
homogeneous background fields in the distant future.
At this stage we expect that the contribution of metric perturbations
can be also important.

\section{Conclusions}
\label{conclusionsec}

In this paper we have studied the dynamics of dark energy in a
Lorentz-violating model with the action given in (\ref{lag}). The model
involves a vector field $B_\mu$ and a scalar field $\Phi$ with mass terms
$M$ and $m$, respectively. The presence of the one-derivative term $\epsilon
\partial_{\mu}\Phi B^{\mu}$ leads to an interesting dynamics at the IR
scales larger than $\epsilon^{-1}$. The phantom equation of state can be
realized without having ghosts, tachyons or superluminal modes in the UV
region.

We have taken into account the contributions of radiation and
non-relativistic matter and studied the cosmological evolution
of the system.
Interestingly, there exists a de~Sitter attractor solution that can be used
for the late-time acceleration. The phantom regime is not an
attractor, 
but we have found that in a range
of parameters, the phantom stage occurs during the transition from the
matter epoch to the final de~Sitter attractor. As is seen, e.g., in
Fig.~\ref{evo1} the equation of state parameter $w$ of dark energy crosses
the cosmological constant boundary towards the phantom region. We
clarified the conditions under which the $w=-1$ crossing is realized
together with the existence of the stable de~Sitter solution.

In the model studied in this paper, the effective Newton's
constant is time-dependent. We have found, however, that this dependence
is typically mild, though for interesting values of parameters it is close
to the experimental bounds.

We have also considered the field perturbations in  Minkowski spacetime
and obtained the momentum scales of instabilities present in the IR region
($p \lesssim \epsilon$). We have found that either tachyons or ghosts
appear for the spatial momenta $p$ smaller than
$\sqrt{(\epsilon^2-M^2)/\alpha-m^2}$, while in the UV region there are no
unhealthy modes. In the cosmological context the presence of tachyons at
the IR scales leads to the amplification of large-scale field
perturbations whose wavelengths are roughly comparable to the present
Hubble radius. There are two tachyonic regions of spatial momenta in this
model: (a) one is sub-horizon and its momenta are characterized by
$M^2/\gamma<p^2<(\epsilon^2-M^2)/\alpha-m^2$;  (b) another is
super-horizon and has $0<p^2<m^2M^2/\epsilon^2$. In the region (a) we
derived the condition under which the perturbations always remain smaller
than the homogenous fields, see Eq.~(\ref{notexceed}). While the existence
of the phantom phase requires that $\epsilon > M$, the condition
(\ref{notexceed}) shows that $\epsilon$ cannot be very much larger than
$M$. Thus the allowed range of $\epsilon$ is constrained to be relatively
narrow. In the tachyonic region (b) the growth of the perturbations is
estimated as $\varphi \propto a^{\alpha m^2/\epsilon^2}$. Since the growth
rate is suppressed by the factor $m^2/\epsilon^2$, this effect is
negligible in the past and at present, though the inhomogeneities can
start to dominate over the homogenous fields in the distant future.

There are several issues  yet to be understood. The presence of the
tachyonic instability on sub-horizon scales may lead to the variation of
the gravitational potential, which can be an additional source of the
late-time integrated Saches-Wolfe effect on the CMB power spectrum.
Another property of this model is the peculiar
time-dependence of the effective Newton's constant, which may result in
interesting phenomenology. 

The model studied in this paper is likely to belong to a wider class of
Lorentz-violating theories exhibiting the phantom behaviour
(see Refs.~\cite{Models} for a number of Lorentz-violating models).
It would be interesting to understand how generic are the features we found in our
particular model --- late-time de~Sitter attractor, transient phantom
stage, time-dependent Newton's constant, sub-horizon
tachyons, super-horizon ghosts, etc. One more direction is to modify our
model in such a way that it would be capable of describing inflationary
epoch rather than the late-time acceleration.
Since it is known that the
spectra of scalar and tensor perturbations produced during the phantom
inflationary phase are typically blue-tilted \cite{Baldi}, this model may
give rise to some distinct features in the CMB spectrum.

\section*{ACKNOWLEDGEMENTS}

This work is supported by RFBR grant (M.\,L.
and V.\,R., 05-02-17363-a), grant of the President of Russian Federation
(M.\,L. and V.\,R., NS-7293.2006.2),
INTAS grant (M.\,L., YSF 04-83-3015), grant of Dynasty Foundation
awarded by the Scientific Board of ICFPM (M.\,L.),
the European Union through the Marie Curie Research 
and Training Network UniverseNet (E.\,P., MRTN-CT-2006-035863)
and JSPS (S.\,T., No.\,30318802).

\section*{\large Appendix A}

We are going to find the effective Newton's constant that
determines the strength of gravitational interactions
at distances shorter than all scales  present in our model,
including $\epsilon^{-1}$, $M^{-1}$, $m^{-1}$ as well as the
Hubble distance. To this end, we neglect the last two terms in the
action (\ref{lag}), and also neglect the time dependence of the
background fields $\phi$ and $X$. We also neglect the space-time
curvature of the Universe, and therefore consider our model in
Minkowski space-time.

Let us impose the gauge $h_{0i}=0$, where $h_{\mu \nu}$ is
the metric perturbation about the Minkowski background.
Then the quadratic Lagrangian for perturbations of metric,
vector and scalar fields is readily calculated,
\begin{eqnarray}
  L &=& \frac{1}{2} \alpha
\left[ \left(\dot{b_i} + \frac{1}{2} X \partial_i h_{00} \right)^2
- \left(\partial_i b_j -  \frac{1}{2} X \dot{h}_{ij} \right)^2
\right]
\nonumber \\
 && + \frac{1}{2} \gamma
\left[ \left(\dot{b_0} + \frac{1}{2} X \dot{h}_{00} \right)^2
- \left(\partial_i b_0 + \frac{1}{2} X \dot{h}_{00} \right)^2
\right]
\nonumber \\
&& +  \frac{1}{2} \left[ \dot{\varphi}^2 - (\partial_i \varphi)^2\right]\,.
\end{eqnarray}
where $X^{2}=B_{0}^{2}$ is the background value. 
Clearly the scalar field $\varphi$ decouples in our approximation,
so we will not consider it in what follows.

By varying the quadratic action with respect to $h_{00}$ and
$h_{ij}$, one obtains $(00)$- and $(ij)$-components of the
linearized energy-momentum tensor for perturbations (note that
$T^{\mu \nu} = - 2 \delta S/ \delta h_{\mu \nu}$). Specifying further
to scalar perturbations with $b_i = \partial_i b_L$ and choosing
conformal Newtonian gauge, $h_{00}= 2 \Phi$, $h_{ij}=
- 2\Psi \delta_{ij}$, one obtains (we keep the standard notation
for the Newtonian potential, even though the same notation was used
for the original scalar field in the main text)
\begin{eqnarray}
& & T_0^0 =\alpha X (X \Delta \Phi - \Delta \dot{b}_L)
+ \gamma X \kern1pt\vbox{\hrule height
1.2pt\hbox{\vrule width 1.2pt\hskip 3pt
   \vbox{\vskip 6pt}\hskip 3pt\vrule width 0.6pt}\hrule
height 0.6pt}\kern1pt (b_0 + X \Phi) \; ,
\nonumber \\
& &  T_j^i =\alpha X \partial_i \partial_j \dot{b}_L -
\delta_{ij} \alpha X^2 \ddot{\Psi}\,,
\end{eqnarray}
where $\Delta = \partial_i \partial_i$ and $\kern1pt\vbox{\hrule height
1.2pt\hbox{\vrule width 1.2pt\hskip 3pt
   \vbox{\vskip 6pt}\hskip 3pt\vrule width 0.6pt}\hrule
height 0.6pt}\kern1pt = \partial_0^2 -\Delta$.
The field equations for $b_0$ and $b_L$ in the absence of sources
for these fields read
\begin{eqnarray}
 && \kern1pt\vbox{\hrule height
1.2pt\hbox{\vrule width 1.2pt\hskip 3pt
   \vbox{\vskip 6pt}\hskip 3pt\vrule width 0.6pt}\hrule
height 0.6pt}\kern1pt (b_0 + X \Phi) =0 \; ,
\nonumber \\
 && - \kern1pt\vbox{\hrule height
1.2pt\hbox{\vrule width 1.2pt\hskip 3pt
   \vbox{\vskip 6pt}\hskip 3pt\vrule width 0.6pt}\hrule
height 0.6pt}\kern1pt
b_L + X (\dot{\Phi} - \dot{\Psi}) = 0\,.
\end{eqnarray}
Now, the longitudinal (proportional to $\partial_i \partial_j$)
part of the $(ij)$-component of the Einstein equations,
in the absence of external anisotropic stresses, gives
\begin{eqnarray}
\Phi + \Psi = 8\pi G \alpha X \dot{b}_L\,,
\end{eqnarray}
while the trace part and $(00)$-component are
\begin{eqnarray}
\ddot{\Psi} + \frac{1}{2} \Delta (\Phi + \Psi)
= - 4\pi G \alpha X^2 \ddot{\Psi} - 4 \pi G p_{ext} \; ,\\
- \Delta \Psi = 4\pi G (\alpha X^2 \Delta \Phi - \alpha X \Delta \dot{b_L})
+ 4 \pi G \rho_{ext} \; ,
\end{eqnarray}
where $\rho_{\rm ext}$ and $p_{\rm ext}$ are energy density and pressure
of an external source.

For time-independent, pressureless source it is consistent
to take all perturbations independent of time and set $b_L = 0$.
Then one finds, as usual, $\Psi = - \Phi$ and obtains the
following equation for the Newtonian potential,
\begin{eqnarray}
(1 - 4\pi G \alpha X^2) \Delta \Phi = 4 \pi G \rho_{ext}\,.
\end{eqnarray}
Thus, the effective Newton's constant in the background field
$X$ is
\begin{eqnarray}
G_* = G (1- 4\pi G \alpha X^2)^{-1}\,.
\end{eqnarray}
This means that the effective Planck mass entering the
Newton's law is given by (\ref{newton}).

\section*{\large Appendix B}
\label{appB}
\subsection{Autonomous equations}

Let us define the following dimensionless variables which are convenient
for studying the dynamical system~\cite{CLW,CST}:
\begin{eqnarray}
x_1=\frac{\kappa \sqrt{\gamma}\dot{X}} {\sqrt{6}H},\quad x_2=\frac{\kappa
\dot{\phi}}{\sqrt{6}H},\quad x_3=\frac{\kappa m \phi}{\sqrt{6}H},\quad
x_4=\sqrt{\frac{4\pi}{3}}\frac{X}{m_{\rm pl}},\quad
x_5=\frac{M}{H}\,,\quad x_6=\frac{\kappa \sqrt{\rho_r}}{\sqrt{3}H}\,.
\end{eqnarray}
Then we obtain the following autonomous equations
\begin{eqnarray}
\label{ba1}
& & x_1'= -3x_1-\frac{3\alpha}{\sqrt{\gamma}}x_4+
\frac{\epsilon}{M}\frac{1}{\sqrt{\gamma}}x_2x_5
+\frac{1}{\sqrt{\gamma}}x_4x_5^2-x_1 \frac{H'}{H}\,, \\
& & x_2'=-3x_2 -\frac{\epsilon}{M} \frac{1}{\sqrt{\gamma}}
x_1x_5-3\frac{\epsilon}{M}x_4x_5-\frac{m}{M}x_3x_5 -x_2\frac{H'}{H}\,, \\
& & x_3'=\frac{m}{M}x_2x_5-x_3\frac{H'}{H}\,, \\
& & x_4'=\frac{1}{\sqrt{\gamma}}x_1\,, \\
\label{ba5}
& & x_5'=-x_5 \frac{H'}{H}\,, \\
& & x_6'=-2x_6 -x_6 \frac{H'}{H}\,,
\end{eqnarray}
and
\begin{eqnarray}
\nonumber
\frac{H'}{H}=-\frac32 \frac{1+x_1^2+x_2^2-x_3^2+x_4^2(3\alpha+x_5^2)
+x_6^2/3+2(\epsilon/M)x_2x_4x_5 +4(\alpha/\sqrt{\gamma})x_1x_4}{1+3\alpha
x_4^2}\,,
\end{eqnarray}
where  prime denotes the derivative with respect to
\[
N \equiv \ln(a)\,.
\]
Equation (\ref{FR}) gives the constraint
\begin{eqnarray}
\Omega_m \equiv \frac{\kappa^2 \rho_m}{3H^2}= 1-x_1^2-x_2^2-x_3^2+x_4^2
\left( 3\alpha +x_5^2 \right)-x_6^2\,.
\end{eqnarray}

Note that the above equations are invariant under the simultaneous change
of the signs of $\phi$ and $X$. Hence it is not restrictive to study the
case of positive $\phi$. Note also that we  study the case of an expanding
Universe with $H>0$.

\subsection{Fixed points}

By setting $x_i'=0$ one formally finds the following six fixed points:
\begin{itemize}
\item (A)~de~Sitter (i):~~$\left( x_1, x_2, x_3, x_4, x_5, x_6  \right)
=\left(0, 0, \frac{\epsilon}{m}\sqrt{\frac{3} {3\epsilon^2/m^2-2\alpha}},
-\frac{1}{\sqrt{3(3\epsilon^2/m^2-2\alpha)}}, \sqrt{3\alpha}, 0
\right)$\,,
\item (B)~de~Sitter (ii):~~$\left( x_1, x_2, x_3, x_4, x_5, x_6  \right)
=\left(0, 0, {\mathit const}, 0, 0, 0 \right)$\,,
\item (C)~matter:~~$\left( x_1, x_2, x_3, x_4, x_5, x_6 \right) =\left(0,
0, 0, 0, 0, 0 \right)$\,,
\item (D)~radiation:~~$\left( x_1, x_2, x_3, x_4, x_5, x_6 \right)
=\left(0, 0, 0, 0, 0, 1 \right)$\,,
\item (E1)~kinetic point (i):~~$\left( x_1, x_2, x_3, x_4, x_5, x_6
\right) =\left(0, 1, 0, 0, 0, 0 \right)$\,,
\item (E2)~kinetic point (ii):~~$\left( x_1, x_2, x_3, x_4, x_5, x_6
\right) =\left(0, -1, 0, 0, 0, 0 \right)$\,.
\end{itemize}

\vspace{0.3cm}

The fixed point (A) is precisely the de~Sitter solution (\ref{phiv}) that
we discussed in Sec.~\ref{sub:prelim}. We will comment on its stability
shortly.

The point (B) is also in some sense a de~Sitter point. It exists even in
the absence of the field $X$ and satisfies the relation $3H^2=\kappa^2
W(\phi)$. To reach the solution (B), the Hubble parameter needs to
increase towards infinity ($M/H \to 0$), and the field 
$\phi$ needs to diverge as well.

The point (C) corresponds to  matter-dominated era satisfying $\Omega_m=1$
and $w_{\rm eff}=0$, whereas the point (D) describes radiation-dominated
epoch with $\Omega_{r}=1$ and $w_{\rm eff}=1/3$.

The points (E1) and (E2) are kinetic solutions satisfying $\Omega_{\rm
DE}=1$ and $w_{\rm eff}=1$. These solutions are used neither for dark
energy nor for radiation/matter dominated epochs.

A cosmologically viable trajectory starts from the radiation point (D),
connects to the matter solution (C) and finally approaches the de~Sitter
point (A). [Note that the initial data (\ref{**initial}) indeed correspond
to $x_1,x_2,x_3,x_4,x_5 \to 0$ as $t \to 0$.] 
To see that 
this sequence of events is indeed possible, let us study  the stability of the
fixed points
against perturbations. 

Let us consider
linear perturbations $\delta x_i$. By perturbing
Eqs.~(\ref{ba1})-(\ref{ba5}) we obtain
\begin{eqnarray}
\delta x_1'&=& \left(-3-\frac{H'}{H}-c_1 x_1 \right) \delta x_1+\left(
\frac{\epsilon}{M} \frac{1}{\sqrt{\gamma}} x_5-c_2 x_1 \right) \delta x_2
-c_3 x_1 \delta x_3 \nonumber \\
& &+\left( \frac{1}{\sqrt{\gamma}}x_5^2-\frac{3\alpha}
{\sqrt{\gamma}}-c_4x_1 \right) \delta x_4 +\left( \frac{\epsilon}{M}
\frac{1}{\sqrt{\gamma}} x_2+\frac{2}{\sqrt{\gamma}}x_4x_5-c_5 x_1
\right)\delta x_5-c_6 x_1 \delta x_6 \,, \\
\delta x_2' &=&  -\left( \frac{\epsilon}{M} \frac{1}{\sqrt{\gamma}}
x_5+c_1 x_2 \right) \delta x_1 -\left(3+c_2 x_2+\frac{H'}{H} \right)
\delta x_2 -\left( \frac{m}{M}x_5+c_3 x_2 \right) \delta x_3 \nonumber \\
& &-\left( 3\frac{\epsilon}{M}x_5+c_4x_2  \right) \delta x_4 -\left(
\frac{\epsilon}{M}\frac{1}{\sqrt{\gamma}} x_1
+3\frac{\epsilon}{M}x_4+\frac{m}{M}x_3+c_5x_2 \right) \delta x_5-c_6 x_2
\delta x_6\,, \\
\delta x_3'&=&-c_1 x_3 \delta x_1+\left( \frac{m}{M}x_5 -c_2 x_3
\right)\delta x_2-c_3 x_3 \delta x_3 -c_4x_3 \delta x_4+\left(
\frac{m}{M}x_2-c_5 x_3 \right) \delta x_5 -c_6 x_3 \delta x_6\,, \\
\delta x_4'&=&\frac{1}{\sqrt{\gamma}} \delta x_1\,, \\
\label{delx5}
\delta x_5'&=& -c_1 x_5 \delta x_1 -c_2 x_5 \delta x_2 -c_3 x_5 \delta
x_3-c_4 x_5 \delta x_4 -\left(\frac{H'}{H}+c_5 x_5 \right)\delta x_5 -c_6
x_5 \delta x_6\,,\\
\label{delx6}
\delta x_6'&=&  -c_1 x_6 \delta x_1 -c_2 x_6 \delta x_2 -c_3
x_6 \delta x_3 -c_4 x_6 \delta x_4 -c_5x_6 \delta x_5- \left(
2+\frac{H'}{H}+c_6 x_6 \right) \delta x_6\,,
\end{eqnarray}
where $\delta (H'/H)=\sum_{i=1}^6 c_i \delta x_i$ with
\begin{eqnarray}
& &c_1=-\frac{3x_1+6(\alpha/\sqrt{\gamma})x_4} {1+3\alpha
x_4^2}x_4\,,\quad c_2=-\frac{3x_2+3(\epsilon/M)x_4 x_5}{1+3\alpha
x_4^2}\,, \quad c_3=\frac{3x_3}{1+3\alpha x_4^2}\,, \nonumber \\
& &c_4=-\frac{3x_4(3\alpha + x_5^2)+3(\epsilon/M)x_2 x_5+
6(\alpha/\sqrt{\gamma}) x_1}{1+3\alpha x_4^2} -\frac{6\alpha
x_4}{1+3\alpha x_4^2}\frac{H'}{H}\,, \nonumber \\
& &c_5=-\frac{3x_4^2x_5+3(\epsilon/M)x_2 x_4} {1+3\alpha x_4^2}\,,
\quad c_6=-\frac{x_6}{1+3\alpha x_4^2}\,.
\end{eqnarray}

The stability of fixed points can be analyzed by considering eigenvalues
of the $6 \times 6$ matrix ${\cal M}$ for perturbations along the lines of
Ref.~\cite{CLW,CST}.

The stability of the de~Sitter point
(A) is important for having the late-time accelerated epoch. This depends
upon the two ratios $\epsilon/m$ and $M/m$ once the parameters $\alpha$
and $\gamma$ are fixed. When $\alpha=1$ and $\gamma=1/2$, for example, the
parameter range of $\epsilon/m$ is determined by the ratio $M/m$. We find
that the point (A) is a stable attractor if the following conditions hold:
\begin{itemize}
\item (i) When $M/m=0.1$, $\epsilon/m>0.817$,
\item (ii) When $M/m=1$, $\epsilon/m>1.35$,
\item (iii) When $M/m=10$, $\epsilon/m>3.52$,
\end{itemize}
When $\epsilon \gg m$, the stability of the point (A) is ensured
automatically unless the ratio $M/m$ is too much larger than unity. In
view of (\ref{epco}) the case $\epsilon \gg m$ is of particular interest.

For another de~Sitter point (B), the eigenvalues are
\begin{eqnarray}
-3, -3, -\frac32 \pm \frac12 \sqrt{9-\frac{12\alpha}{\gamma}}, 0, -1/2\,,
\end{eqnarray}
This means that this point is marginally stable. The zero eigenvalue comes
from the perturbation equation for $\delta x_5$. If $H$ continues to
increase toward the solution (B), this eigenvalue actually obtains a small
negative value, as can be seen from Eq.~(\ref{delx5}). Thus in such a case
the point (B) is stable. However, we know that the phantom phase is
realized only for finite field values bounded by $\phi_A$, see 
Eq.~(\ref{philessphia}). Hence it
 is not possible that the actual solutions approach the point (B) with
infinite $H$ and $\phi$.

The matter point (C) has  the eigenvalues
\begin{eqnarray}
3/2, -3/2, -\frac43 \pm \sqrt{9-\frac{48\alpha}{\gamma}}, 0, -1/2\,,
\end{eqnarray}
which shows that the matter era corresponds to a saddle point with one
positive eigenvalue. Hence the solutions eventually repel away from this
fixed point even if they temporarily approach it.

The radiation point (D) has the eigenvalues
\begin{eqnarray}
-\frac12 \pm \frac12 \sqrt{1-\frac{12\alpha}{\gamma}}, -1, 0, 2, 1\,.
\end{eqnarray}
One finds that the radiation epoch corresponds to a saddle point with two
positive eigenvalues.

The kinetic points (E1) and (E2) have the eigenvalues
\begin{eqnarray}
3, 3, 1, 0, \sqrt{\frac{3\alpha}{\gamma}}i,
-\sqrt{\frac{3\alpha}{\gamma}}i\,,
\end{eqnarray}
which shows that they are unstable.

The above stability analysis shows that the sequence of radiation, matter
and de~Sitter epochs can indeed be realized.

\bibliographystyle{unsrt}

\end{document}